\documentclass[reprint,amsmath,amssymb,aps]{revtex4-1}
\usepackage{graphicx}
\usepackage{amsmath}
\usepackage{epsfig}
\usepackage[colorlinks]{hyperref}
\usepackage{siunitx} 
\usepackage{ulem} 

\begin{document}

\title{Observation of binary phase states of time-multiplexed degenerate optical parametric oscillator pulses generated using a nonlinear fiber Sagnac loop}
\author{Hsin-Pin Lo}
\email{hsinpin.lo.cn@hco.ntt.co.jp}
\affiliation{NTT Basic Research Laboratories, NTT Corporation, 3-1 Morinosato Wakamiya, Atsugi, Kanagawa, 243-0198, Japan}
\author{Takahiro Inagaki}
\affiliation{NTT Basic Research Laboratories, NTT Corporation, 3-1 Morinosato Wakamiya, Atsugi, Kanagawa, 243-0198, Japan}
\author{Toshimori Honjo}
\affiliation{NTT Basic Research Laboratories, NTT Corporation, 3-1 Morinosato Wakamiya, Atsugi, Kanagawa, 243-0198, Japan}
\author{Hiroki Takesue}
\email{hiroki.takesue.km@hco.ntt.co.jp}
\affiliation{NTT Basic Research Laboratories, NTT Corporation, 3-1 Morinosato Wakamiya, Atsugi, Kanagawa, 243-0198, Japan}


\begin{abstract}
We generated time-multiplexed degenerate optical parametric oscillator (DOPO) pulses using a nonlinear fiber Sagnac loop as a phase-sensitive amplifier (PSA) where the pump and amplified light in pump-signal-idler degenerate four-wave mixing can be spatially separated. By placing the PSA in a fiber cavity, we successfully generated more than 5,000 time-multiplexed DOPO pulses. We confirmed the bifurcation of pulse phases to 0 or $\pi$ relative to the pump phase, which makes them useful for representing Ising spins in an Ising model solver based on coherent optical oscillator networks. We also confirmed inherent randomness of the DOPO phases using the NIST random number test.
\end{abstract}

\maketitle

As various systems and networks in our society become larger and more complex, their optimization is becoming an increasingly important issue. Most such tasks are classified as combinatorial optimization problems, which are in general difficult to solve with conventional digital computers. A promising approach to solving such problems is to use the Ising model \cite{ising}. In this approach, a combinatorial optimization problem is converted into a ground-state search problem of the Ising model \cite{lucus} which is solved by using an artificial spin network created with physical systems, such as superconducting quantum bits \cite{dwave}, trapped ions \cite{ion}, single-electron devices \cite{single}, and nanomechanical oscillators \cite{nano}. A coherent Ising machine (CIM) is one such Ising model solver which uses degenerate optical parametric oscillators (DOPOs) as artificial spins \cite{wang2013,ina2,peter,fabian,roma,baba,okawachi}. A DOPO is produced by placing a phase-sensitive amplifier (PSA) in an optical cavity, which amplifies the 0 or $\pi$ phase component of incoming light through signal-idler degenerate optical parametric amplification \cite{levenson}. Because of the PSA, the phase of a DOPO light takes only 0 or $\pi$, with which an Ising spin state can be expressed by, for example, allocating the phase 0 ($\pi$) as spin up (down). With the use of temporal multiplexing, from four to more than one million DOPO pulse have been generated from a single optical cavity using $\chi_2$ \cite{alireza,ina2,peter} and $\chi_3$ \cite{ina1,takesue} optical nonlinearities. Interactions among the time-multiplexed DOPO pulses have been implemented by direct optical coupling using delay interferometers \cite{alireza,ina1,takata} and a measurement-feedback (MFB) scheme \cite{ina2,peter,ryan}. The coupled DOPOs tend to oscillate in the phase configuration that minimizes the total loss, which corresponds to the ground state of the given Ising model problem. 

It has been demonstrated that a CIM with MFB can find a fair solution to a maximum-cut problem of a 2,000-node fully connected graph faster than simulated annealing on a CPU \cite{ina2}. An important element of the MFB is a balanced homodyne detector (BHD) measurement of DOPO pulses, which requires a local oscillator (LO) light whose phase is locked to that of the DOPO pulses. In addition, the phase of an optical pulse for feedback should also be locked to DOPO pulses. To obtain such phase reference lights, previous MFB-CIM experiments have used a configuration based on a signal-idler degenerate optical parametric amplifier pumped by a light generated via second-harmonic generation (SHG) using $\chi_2$ nonlinearity in two periodically-poled lithium niobate (PPLN) waveguides \cite{ina2,peter}. In those experiments, the seed light for the SHG was used as a phase reference to the squeezed vacuum light generated from the degenerate optical parametric amplifier. On the other hand, in the $\chi_3$ based DOPO generation, two pumps with different wavelengths were employed for signal-idler degenerate optical parametric amplification based on four-wave mixing (FWM) so that the pumps and the degenerate signal could be spectrally separated \cite{ina1,takesue}. Since the pump and degenerate signal wavelengths are different in this configuration, the pump lights cannot be used as an LO, and thus BHD measurement could not be performed in those experiments.
As a result, an MFB-CIM has not been realized using $\chi_3$ nonlinearity. 

\begin{figure*}
\includegraphics[width=15cm]{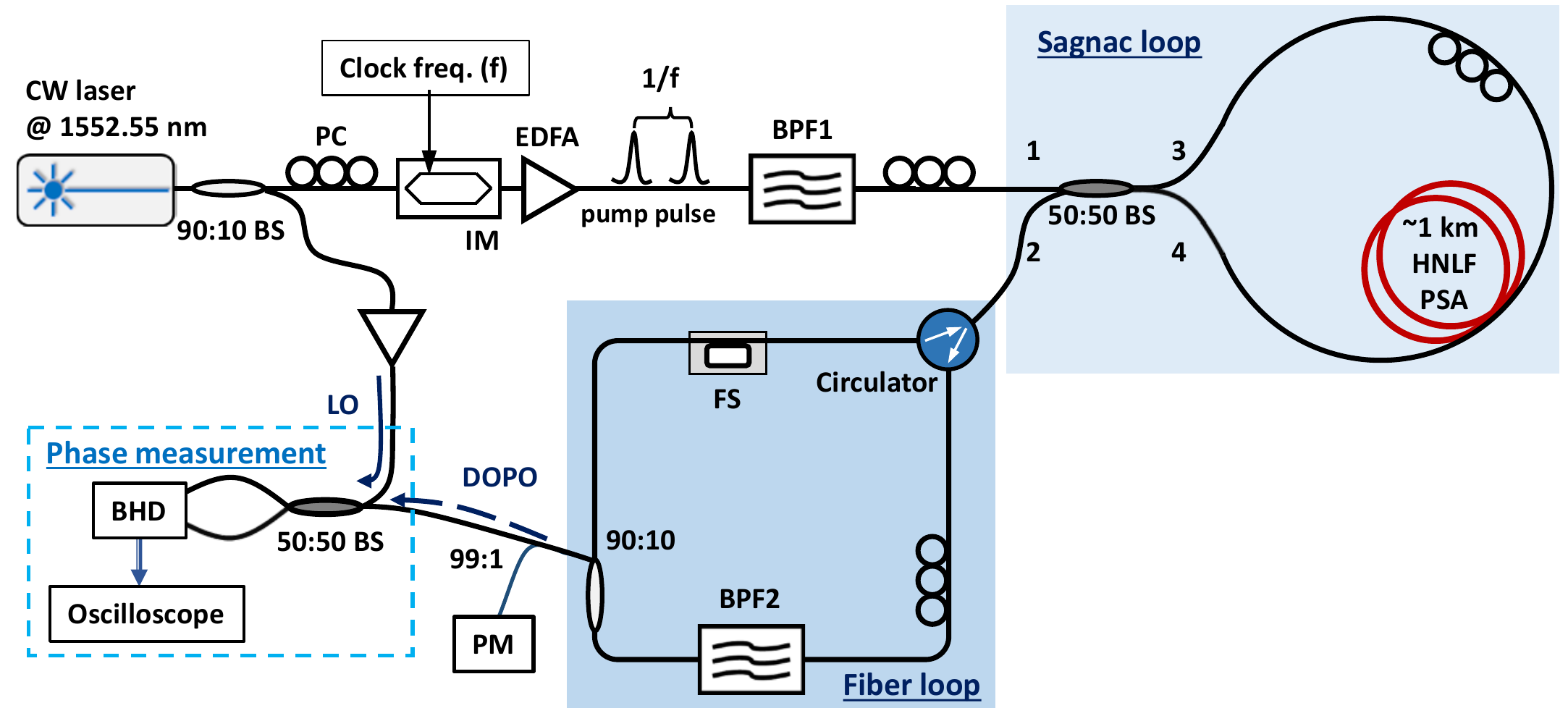}
\caption{Experimental setup for DOPO pulses generation through degenerated FWM based on a Sagnac interferometer. BS, beam splitter; PC, polarization controller; IM, intensity modulator; EDFA, erbium-doped fiber amplifier; BPF, band pass filter; HNLF, highly nonlinear fiber; FS, fiber stretcher; PM, power meter; LO, local oscillator; BHD, balanced homodyne detector.}
\end{figure*}

In this paper, we report the generation of time-multiplexed DOPO pulses based on pump-signal-idler degenerate FWM. To separate the pump and degenerate signal lights in all-degenerate FWM, we employed a nonlinear Sagnac loop configuration \cite{imajuku,kumar,Levandovsky}, which enabled us generate more than 5,000 time-multiplexed DOPO pulses, and at the same time, we can use the pump as a local oscillator for a BHD measurement. Note that a DOPO based on all-degenerate FWM in a nonlinear Sagnac loop has already been reported by Serkland et al. \cite{kumar}, who demonstrated the oscillation by showing the optical spectrum, performing autocorrelation measurements, and analyzing the cross correlation between the output-signal pulses and the compressed laser pulses. However, the coherence properties of the output light was not investigated. Our work complements their pioneering work by confirming the bifurcation of DOPO phases to 0 and $\pi$, which is their most important characteristic as artificial spins for the CIM. Using the random number test suite developed by National Institute of Standards and Technology (NIST) \cite{nist}, we also demonstrate that the discretized DOPO phases constitute an unbiased random number.

The experimental setup is shown in Fig.~1. A continuous-wave (CW) laser with a wavelength of 1552.55 nm was modulated into a pulse train at a repetition frequency (f) of 1 GHz by an optical intensity modulator (IM). The pulse train was then amplified by an erbium-doped fiber amplifier (EDFA) followed by an optical bandpass filter (BPF1) with a 0.8-nm bandwidth. The pump pulses were then input into a Sagnac loop composed of a 50:50 beam splitter (BS), a highly nonlinear fiber (HNLF), and a polarization controller (PC). The length, nonlinear coefficient, zero-dispersion wavelength, and dispersion slope of the HNLF are 1 km, 21 W$^{-1}$  km$^{-1}$, 1537 nm, and 0.03 nm$^{-2}$  km$^{-1}$, respectively.

\begin{figure}
\includegraphics[width=7.5cm]{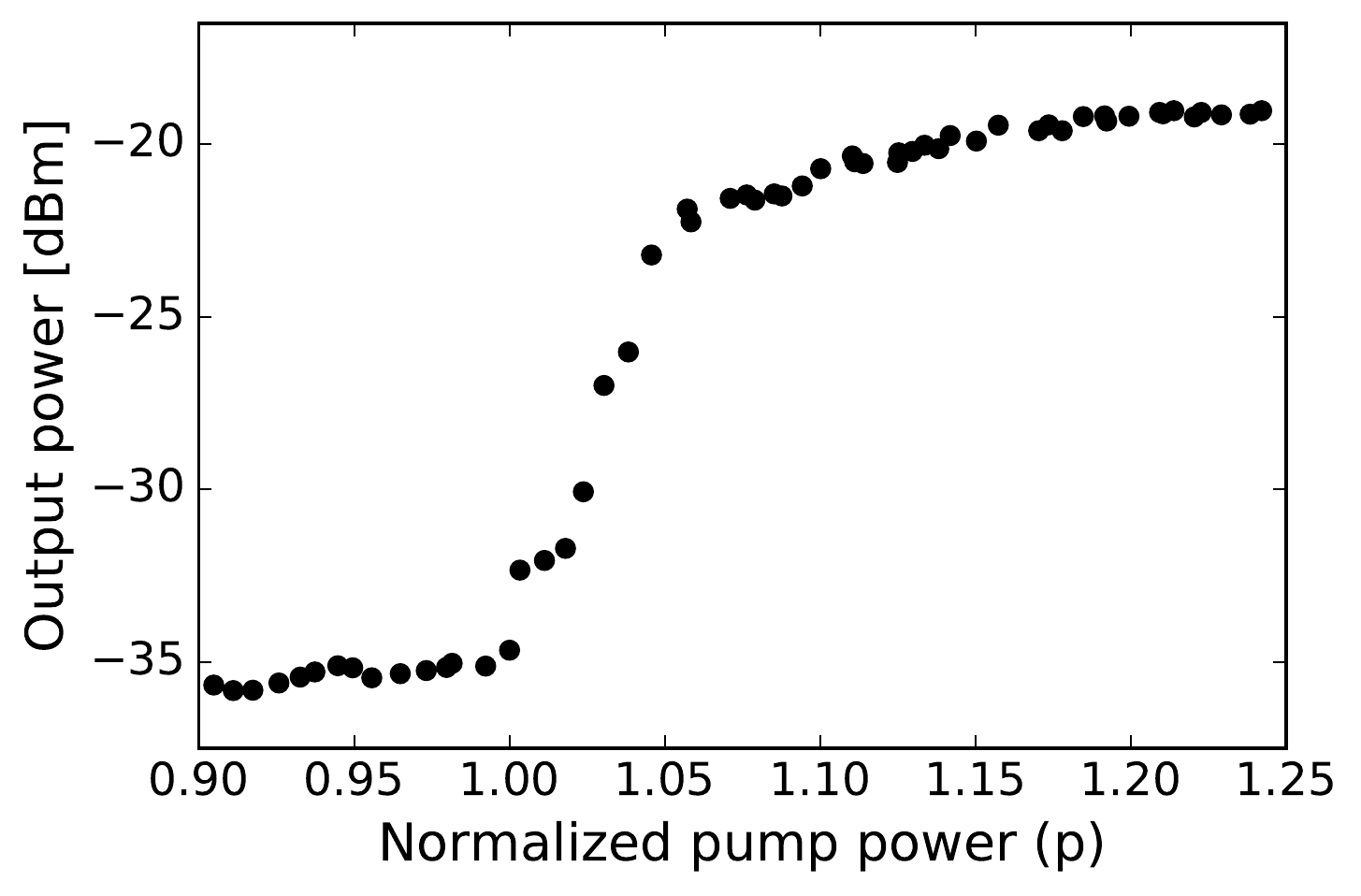}
\caption{DOPO output power as a function of normalized input pump amplitude ($p$).}
\end{figure}

The pump pulse input from port 1 of the 50:50 BS was split into two directions to ports 3 and 4 and launched into both ends of the HNLF, while a vacuum state was input into port 2. At the HNLF, squeezed vacuum pulses \cite{Yu, Meissner} were generated via degenerate FWM in both clockwise and anticlockwise propagating directions. Because of the constructive interference at the respective ports in a nonlinear Sagnac loop, the pump pulse was output from port 1, while a squeezed vacuum state was output from port 2 \cite{Fermann}. The HNLF was not a polarization-maintaining fiber, so we adjusted the PC in the Sagnac loop to achieve good spatial separation of the pump and squeezed lights. As a result, we obtained a $\sim$30 dB pump suppression at port 2. 
In this way, we were able to achieve squeezed vacuum pulse generation as well as degenerate optical parametric amplification based on all-degenerate FWM with a HLNF. 
The squeezed vacuum pulses from port 2 were then input through a circulator into a fiber loop, which forms an optical cavity with the Sagnac loop. The fiber loop included a PC, a BPF (BPF2) with a 0.2-nm bandwidth, and a fiber stretcher (FS), which was used to lock the cavity phase so that the output light from the cavity was maximized. 
Consequently, a quadrature of the amplitude of a circulating light chosen by the PSA was always amplified and the conjugate one deamplified.
After passing through the fiber loop, the squeezed vacuum pulses output from the circulator were again launched into the Sagnac loop from port 2, where they were amplified with a PSA. 
With repetitive phase-sensitive amplification in a cavity, the amplified squeezed vacuum pulse undergoes spontaneous symmetry breaking if the PSA gain exceeds the oscillation threshold, resulting in the oscillation in either the 0 or $\pi$ phase \cite{alireza}. Also, it is known that we can observe superposition of coherent states if the PSA gain is large and the cavity loss is very small \cite{wandc}. 
We inserted a 90:10 fiber coupler to extract 10\% of DOPO signal from the cavity for output power measurement and phase measurement using a BHD. A portion of light extracted from the output of the CW laser was used as an LO for BHD measurement.
The round-trip time of the fiber cavity, including the nonlinear Sagnac loop was $\sim$5.555 $\mu$s, which is dominated by the delay time of the 1-km HNLF. Thus, by pumping the PSA at a repetition frequency of 1 GHz, we generated 5,555 independent DOPO pulses.

\begin{figure}
\includegraphics[width=7.5cm]{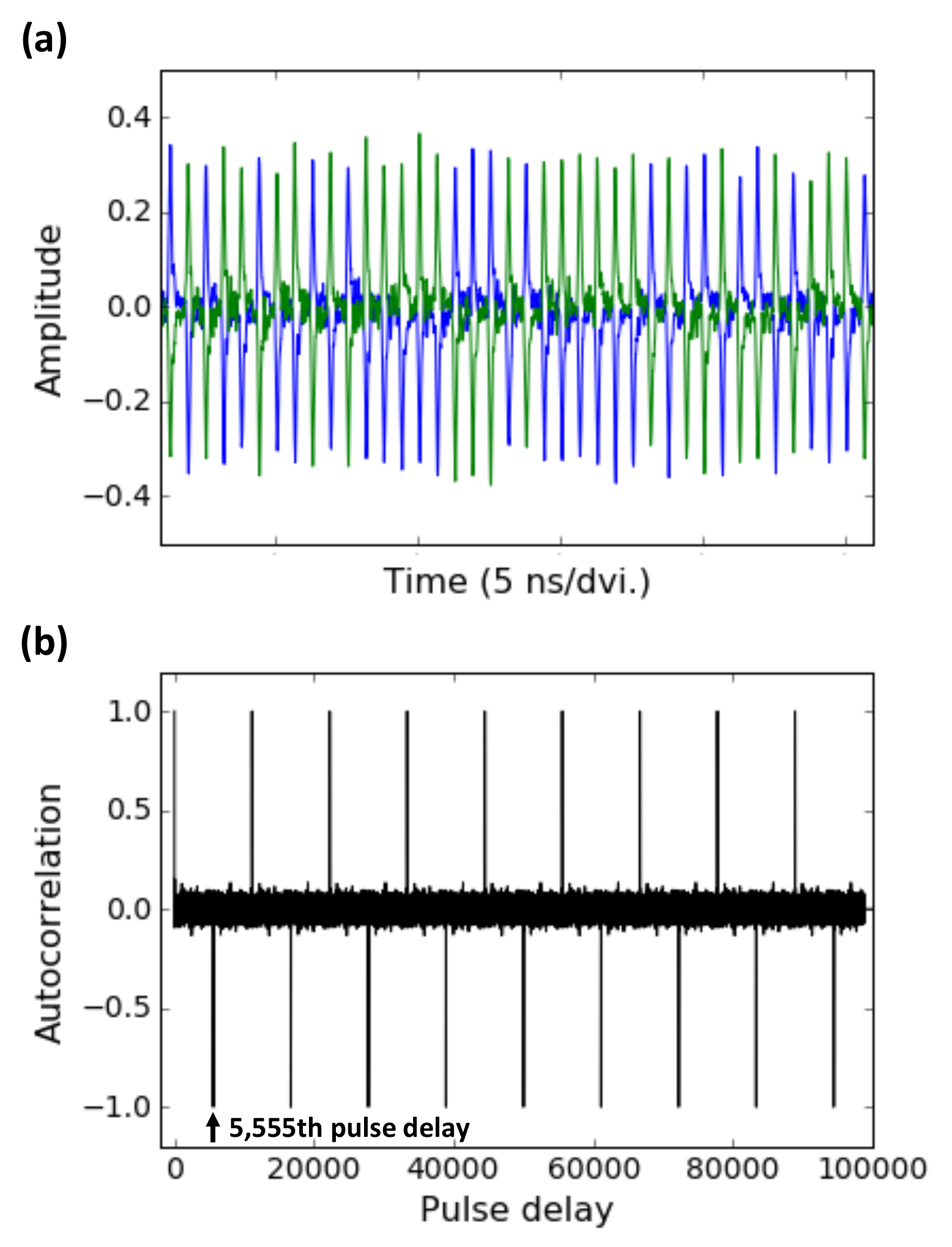}
\caption{(a) Result of BHD measurement of DOPO pulses. The green curve shows a waveform with its temporal position shifted by 5,555 pulses. (b) Autocorrelation of BHD measurement result in (a) as a function of pulse delay.}
\end{figure}

First, we performed the DOPO output power measurement using a small portion of DOPO lights extracted with couplers while changing the input pump power as a function of the normalized pump amplitude ($p$). The result is shown in Fig. 2, where clear threshold behavior was observed at an estimated peak pump power of  $\sim$11.7 mW at $p=1$. 

Then, we performed the BHD measurement to investigate the coherence property of a DOPO pulse sequence. We mixed the DOPO pulses and LO on a 50:50 BS and then connected both output ports to a BHD to measure the quadrature amplitudes of the DOPO pulses using an oscilloscope. Figure~3(a) shows the result, which indicates that the absolute values of amplitudes were almost the same. Their signs take plus or minus randomly and the ratio plus pulses to minus ones was 1:0.99986, which indicated that the probability of pulse generation of 0 or $\pi$ phase was the same. Thus, we could confirm that the pulse phase values were limited to only 0 or $\pi$. 

Although the DOPO phase is random, once the DOPO stably oscillates at above the threshold level, the phase pattern of the DOPO pulses should be preserved. To confirm this, we took the autocorrelation of the BHD measurement result as shown in Fig.~3(b), which is plotted as a function of pulse delay in Fig.~3(a). We observed a nearly perfect negative correlation at every 5,555th pulse seventeen times, which implies that the DOPO pulses oscillated stably at least for seventeen circulations in the fiber cavity. The negative correlation indicates that the DOPO pulses underwent a $(2n + 1)\pi$ ($n$: integer) phase shift per circulation in the cavity. This experiment confirmed that we could generate 5,555 independent DOPO pulses that can be used as artificial spins for a CIM. 

We also confirmed the randomness of the DOPO phase using the NIST random number test (Special Publication 800-22 Statistical Test Suite) \cite{nist}. Because of the instability of the Sagnac loop (explained later), long-term operation of the DOPO was not possible. Consequently, the number of random bits obtained from the BHD measurement of the DOPO pulses was limited, which was too short for some of the NIST tests. Therefore, we performed eight tests for which the number of the obtained random bit suffices. The results are summarized in Table~\ref{table}, which indicate that the obtained random bits passed all eight tests with the available number of bits. This suggests that we can generate photonic artificial spins with inherent randomness with the present setup.   

In the present experiment, the success of DOPO pulse generation was strongly dependent on the pump suppression ratio at port 2. When the ratio was in sufficient, the pump leakage to the fiber cavity became a seed light for the parametric oscillation. Consequently, all the pulses oscillated at the same phase, losing their inherent phase randomness. Although we could successfully suppress the pump leakage with a careful adjustment of PCs in the Sagnac loop and fiber loop, we observed a temporal change in the pump suppression ratio due to the temperature change in the HNLF, which made long-term operation of this setup difficult. We should be able to overcome such instability by using a polarization-maintaining HNLF \cite{pmhnlf}. 

An advantage of an HNLF over PPLN waveguides in DOPO generation is that a highly efficient PSA can be achieved relatively easily with an HNLF. Although high-gain PSAs based on PPLN waveguides have been reported \cite{umeki}, such waveguides require highly precise fabrication technologies to reduce waveguide loss, which results in high cost and low yields. On the other hand, in a PSA based on an HNLF, it is relatively easy to increase the effective interaction length by simply increasing the fiber length, because of the small loss. In fact, a very large optical parametric amplification gain has been reported using an HNLF \cite{gain}. Thus, the DOPO based on the Sagnac loop including an HNLF will be a promising candidate as an artificial spin system for a MFB-CIM. 

An important consideration in DOPO generation using an HNLF is to suppress the phase-insensitive amplification (PIA) that accompany PSA. In the HNLF, we should observe not only PSA based on all-degenerate FWM but also PIA caused by pump-degenerate, signal-idler non-degenerate FWM. However, the noise lights generated by PIA had optical frequencies that are different from the frequency of the degenerate signal from the PSA. As a result, the noise from the PIA process was suppressed by the bandpass filter, which only passed the degenerate signal. Another PIA process that can possibly occur in the  HNLF is spontaneous Raman scattering (SRS). However, since the peak of the Raman gain in an optical fiber is about 100 nm from the pump wavelength \cite{agrawal}, the noise from SRS can again be removed by the bandpass filter. Thus, we consider that the effect of the PIA noise was well suppressed, and as a result, we could observe the phase bifurcation, which indicates PSA is the dominant amplification process in our system.

In conclusion, we generated DOPO pulses through a degenerate FWM based on a Sagnac interferometer and studied their coherence properties. We confirmed that the phases of our DOPO pulses were bifurcated to 0 or $\pi$. We observed a nearly perfect correlation at every 5,555th DOPO pulses, which implies that our DOPO pulses were operated stably in a fiber cavity. Furthermore, the random numbers generated from the phase measurement of DOPO pulse sequence passed the NIST random number test. There results indicate that our DOPO pulses could be useful for simulating Ising spins.

\begin{table}[h!]
\caption{Results of NIST Special Publication 800-22 statistical tests.
Our sample consists of 10 sets of 555 bits. The P-value indicates the probability that a perfect random number would be generated that is less random than the test sequence \cite{nist}. The proportion indicates the ratio of sequences that pass the tests with a level of significance $\alpha$ $>$ 0.01, which indicates the upper bound for the probability of incorrectly rejecting the null hypothesis. In our system, the minimum proportion is 8/10.}
\label{table}
\centering
\begin{tabular}{lccc}
\hline 
Statistical test & P-value & Proportion & Result\\ 
\hline
Frequency & 0.534146 & 10/10 & Pass\\ 
BlockFrequency & 0.350485 & 10/10 & Pass  \\
CumulativeSums & 0.739918 & 10/10 & Pass  \\
Runs & 0.008879 & 9/10 & Pass \\
LongestRun & 0.534146 & 10/10 & Pass \\ 
FFT & 0.122325 & 10/10 & Pass \\
ApproximateEntropy & 0.017912 & 9/10 & Pass\\
Serial & 0.739918 & 10/10 & Pass \\
\hline 
\end{tabular}
\end{table}




\end{document}